\documentclass{egpubl}
\usepackage{ICAT-EGVE2014}
\usepackage{hyperref}
\usepackage{fancyhdr}

 \WsPaperJoint         
 \electronicVersion 

\ifpdf
 \usepackage[pdftex]{graphicx}
 \DeclareGraphicsExtensions{.eps}
\else \usepackage[dvips]{graphicx} \fi

\usepackage{t1enc,dfadobe}

\usepackage{egweblnk}
\usepackage{cite}

\title[Interpretation of Tactile Sensation using an AFM Interface to Operate a Virtual Avatar]%
{Interpretation of Tactile Sensation using an Anthropomorphic Finger Motion Interface to Operate a Virtual Avatar}

\author[Y. Ujitoko \& K. Hirota]
       {Y.\,Ujitoko$^{1}$
        and K. Hirota$^{2}$
        \\
         $^{1,2}$The University of Tokyo, Japan
       }

\begin{document}

%
%

\maketitle

\thispagestyle{fancy}
\fancyhf{}
\chead{This is a preprint of an Article accepted for publication in ICAT-EGVE 2014 - International Conference on Artificial Reality and Telexistence and Eurographics Symposium on Virtual Environments \copyright 2014 The Eurographics Association.}

\begin{abstract}
 {
The objective of the system presented in this paper is to give users tactile feedback while walking in a virtual world through an anthropomorphic finger motion interface.
We determined that the synchrony between the first person perspective and proprioceptive information together with the motor activity of the user's fingers are able to induce an illusionary feeling that is equivalent to the sense of ownership of the invisible avatar's legs.
Under this condition, the perception of the ground under the virtual avatar's foot is felt through the user's fingertip.
The experiments indicated that using our method the scale of the tactile perception of the texture roughness was extended and that the enlargement ratio was proportional to the avatar's body (foot) size.
In order to display the target tactile perception to the users, we have to control only the virtual avatar's body (foot) size and the roughness of the tactile texture.
Our results suggest that in terms of tactile perception fingers can be a replacement for legs in locomotion interfaces.
}
\end{abstract}


\section{Introduction}

Walking in the virtual world is a fundamental task that virtual reality technologies should be able to realize.
Providing the ability to walk through large-scale virtual scenes is of great importance for many applications such as training, architectural visits, tourism, entertainment, games, and rehabilitation.
Over the years, a large number of technical approaches have been proposed and investigated.
Most of these addressed the locomotion interface that supports human leg motions themselves\cite{darken1997omni}\cite{iwata1999path}.
An advantage of this approach is that it facilitates tactile or force feedback directly to the user's soles.
The design of haptic stimulations that correspond to the rendering of floor attributes and ground properties is a key for natural locomotion simulation.
However, the devices necessary to support full-body motions tend to be large and complicated.
Another method for operating a virtual avatar is to use fingers as input instead of legs.
For example, finger motions that pantomime leg movements can be used as input for operating bipedal walking robots\cite{fernando2009operating} and for navigation\cite{kim2008finger}.
In this method, the physical motion is scaled down, and consequently physical body fatigue should be diminished, as compared with that incurred by the full-body locomotion technique.
To the best of our knowledge, no study has been conducted that focuses on the tactile feedback involved when a user controls a virtual avatar using finger motions.
An essential factor of locomotion using legs is the sense of contact between the feet and the ground surface.
We aim to  explore the possibility of giving users this sense of contact adequately by using finger movements and passive tactile feedback.
Two questions arise concerning the presentation of tactile sensation to the fingers: can the tactile stimulation on the user's fingers be felt as the tactile experience on the sole of a virtual avatar, and does the difference in size between fingers and legs affect tactile perception?
In this paper, we address hypotheses for answering the questions above and describe the experiments that we conducted.
The preliminary results showed that illusory ownership of the virtual avatar was evoked.
To the best of our knowledge, this is the first paper that addresses the sense of ownership of an invisible virtual avatar's legs that is induced by stimulating the user's fingers.
This paper also refers for the first time to a change in tactile perception being induced by using the fingers to mimic the motion of the avatar's legs.
We would like to apply this knowledge to designing a locomotion interface and confirm that fingers can be substituted for legs in terms of tactile sensation.

\section{Related work}

Multiple locomotion interfaces have been proposed and invented, such as gamepads and treadmills.
However, no natural, general-purpose locomotion interface exists.
What is most needed for locomotion in virtual environments (VEs)?
Body-based information about the translational and rotational components of movement helps users to perform a navigational search task\cite{Ruddle2009}.
Waller et al.\cite{waller2004body} examined the body-based cues resulting from active movements that facilitate the acquisition of spatial knowledge.
Full body locomotion is possible on walking simulators.
For example, the Omnidirectional treadmill\ \cite{darken1997omni} or Torus treadmill\cite{iwata1999path} enable users to move virtually in any direction while their position in the real world is fixed.
An advantage of this approach, which supports full-body motion, is that it facilitates tactile or force feedback directly to the user's soles.
The haptic experience that corresponds to the rendering of floor attributes and ground properties is a key to natural locomotion.
However, entire devices supporting full-body motions tend to be large and complicated.

Instead of simulating full-body locomotion, several interaction techniques using a full-body metaphor were presented.
The Walking-in-Place technique\cite{slater1995taking} avoids the user being relocated in the real world.
The user walks by utilizing walk-like gestures instead of actually walking.
The Step WIM technique\cite{laviola2001hands} allows the user to interact with VEs through a hand-held miniature copy of the scene.
Other techniques that transform the VE or user's motion by rotation or scaling include redirected walking\cite{razza}, scaled-translational-gain\cite{williams2006updating}\cite{williams2006exploring}, seven-league-boots\cite{interrante2007seven}, and motion compression\cite{nitzsche2004motion}.
Generally, these metaphor techniques are lacking in terms of kinesthetic feedback.

Another method for realizing locomotion metaphor is to use fingers instead of legs.
It takes advantage of the structural similarities between fingers and legs.
Users operate their fingers as if the fingers are walking or running.
For example, two fingers were used to mimick leg movements for generating full-body animations\cite{lockwood2012fingerwalking}.
In this study, the performance of "finger walking" on a touch-sensitive tabletop was recorded and properties were analyzed to determine which motion parameters are most reliable and expressive for generating full-body animations.
It was concluded that finger walking is a naturally chosen and comfortable performance method.
The finger walking in place (FWIP) method was proposed in \cite{kim2008finger}\cite{kim2010effects}.
In terms of spatial knowledge acquisition, FWIP results showed a better performance than did a rate-based translation and turning system (i.e., Joystick) during maze navigation tasks.
This technique can also be used to provide input for operating bipedal walking robots\cite{fernando2009operating}.
The proposed system detected finger movements and recreated synchronous motions in the bipedal robot.
This approach has a key advantage in its smaller scale.
The physical motion required by this approach diminishes the body fatigue as compared with the physical locomotion technique by virtue of the difference in scale\cite{kim2010effects}.
The device or enviroments around the fingers that give fingers tactile stimulations can all be scaled down as compared with full body locomotion devices.
However, the sense of touch was not mentioned in these studies.

Tactile or force feedback would further extend the VR-based walking experience.
Therefore, the issue addressed is the design of haptic stimulations.
The following devices that stimulate the user's soles were proposed in previous studies.
Haptic shoes that present vibrotactile stimulations have been demonstrated in an interactive context.
The early fantastic phantom slipper system enabled the user to virtually step on vibrating objects\cite{kume1998foot}.
FootIO\cite{rovers2005footio} can be used for remote communication enhanced with tactile feedback at the level of the feet.
Haptic floors correspond mostly to actuated ground surfaces made of individually activated tiles.
The tiles of the ATR Alive Floor\cite{sugihara1998terrain} are triangular plates that are locally displaced and oriented to approximate the local shape of the virtual simulated terrain.
The EcoTiles\cite{visellarchitectural} are equipped with large loudspeakers that generate a high-amplitude vibratory feedback under the feet.
Auditory feedback can also be embedded in and emitted by shoes.
Sonic shoes were also proposed in\cite{papetti2010audio}.
In these systems, the devices tend to be large and complicated.

The objective of our study is to allow users to feel a sense of touch on the avatar's sole through stimulation of their fingers.
It is necessary for users to interpret their fingers as the avatar's legs.
In other words, the sense of ownership of the avatar's legs is required.
The flexibility of ownership has been investigated experimentally through the illusory embodiment of a rubber limb.
In the Rubber Hand Illusion (RHI)\cite{botvinick1998rubber}, synchronous touches are applied to a rubber hand in full view of the participant and the real hand, which is hidden behind a screen, produce the sensation that the touch originates from the rubber hand and a feeling of ownership of the artificial hand.
This suggests that the temporal and spatial patterns of visual and somatosensory signals play an important role in how we come to experience that a limb is part of our own body.
It has also been shown that the illusion can be induced with visuomotor stimulation, meaning that the virtual hand moves in synchronization with the movements of the corresponding hidden real hand\cite{sanchez2010virtual}\cite{dummer2009movement}.
In other words, the RHI can be induced by two methods: visuotactile and visuomotor stimulation.
Kokkinara et al.\cite{kokkinara2014measuring} showed that visuomotor synchronous stimulation contributes more to the attainment of the illusion than does visuotactile stimulation.
Maravita \cite{maravita2004tools} reported that the illusion can be created with objects the shape of which is not similar to that of the hand.
Bruno et al.\cite{bruno2010haptic} found that synchronous visuotactile stimulation led participants to report that an enlarged or reduced hand replica had become their own hand.
They also revealed that multisensory stimulation can alter the internal representation of one's hand, affecting the interpretation of active touch.
As a result, participants judged an actively felt object to be larger (after exposure to the enlarged hand) or smaller (after exposure to the reduced hand) than a standard object of identical size felt by the other hand.
Similar results were obtained through the size-weight illusion\cite{linkenauger2011body}.
The size-weight illusion is a well-known phenomenon that occurs when people lift two objects of equal weight but differing sizes and perceive that the larger object feels lighter.
Therefore, if the apparent hand size influences the perceived object size, it should also influence the object's perceived weight.
The results suggested that an increase in hand size made the subsequent objects feel small, evoking a size-weight illusion.

According to our previous knowledge, it may be expected that, when virtual avatar's legs move synchronously with the movements of the corresponding user's fingers, the sense of ownership will occur.
Further, it should be possible to change the haptic perception of the ground according to the avatar's foot size.

\section{Interpretation of the tactile sensation}

\subsection{Interface using anthropomorphic finger motions}

\begin{figure}[htb]
  \centering
  \includegraphics[width=.95\linewidth]{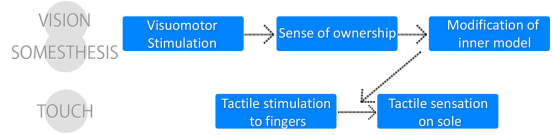}
  %
  %
  \caption{\label{fig:1}
           Processing chart.}
\end{figure}

In our study, we call the following settings an anthropomorphic finger motion interface and conducted experiments using that interface.
Finger movements mimicking actual walking (not a walking in place technique) simulate an avatar's leg movements.
The first person perspective of the avatar synchronizes with finger motions and is displayed to the user.
In this study, the user was not allowed to turn around or change the direction of the avatar.
Users were able to move back and forth and around while the first person view remained in the forward direction.
In order to create the sense of touch while walking in the virtual world using this interface, the illusory ownership of the avatar's legs is required.
According to the findings described in the last section, we speculate that the visuomotor synchronous stimulation to the fingers can cause the illusion.
This corresponds to the first arrow in the Figure~\ref{fig:1}.

As compared with previous studies on illusory ownership, the current study addresses two new points.
First, the body part that is stimulated and the body part to which the stimulation is transferred are different in this study.
We deal with whether illusory ownership of the avatar's legs is induced by the user's finger movement.
In constrast, in the previous studies, the stimulated and target body parts were the same.
Second, in this study the target body part (avatar's feet) is invisible to the user.
This is because the first person perspective usually does not provide visual information of the legs while walking or running in the real world.
In contrast, in the previous studies, the dummy hand or legs in the virtual or real world were visible.
Despite these two points of difference, we presumed that the illusory ownership could be generated as in a previous study\cite{kokkinara2014measuring}.
If the sense of ownership can be achieved, it will cause users to interpret the haptic stimulation through their fingers as deriving from the avatar's legs.
Then, attention should be paid to the difference in scale between fingers and legs.
A representation of one's own body is implicitly used to calibrate the perception of external objects (Figure~\ref{fig:1}).
According to the studies presented in \cite{haggard2009rubber}\cite{bruno2010haptic}, we suggest that haptic perception of the ground while walking in the virtual world can be controlled by varying the target body size.

\subsection{Hypothesis}

For the sense of touch, we focus on the texture roughness.
Our hypothesis predicts that if a user is absolutely convinced that the tactile sensation experienced by his/her fingers is the same as that experienced by the legs, the scale of the tactile perception of the texture feature size can be extended to one that is equivalent to the sole of the foot.
For example, when their fingers move on gravel, users feel that they are moving their legs on large rocks in the VEs.
Further, the tactile perception in the VEs can be changed by controlling the avatar’s sole size.
If the avatar’s sole is larger, the user’s tactile perception will be that objects in the VEs are larger proportionally, and vice versa.

\begin{figure}[htb]
  \centering
  \includegraphics[width=.95\linewidth]{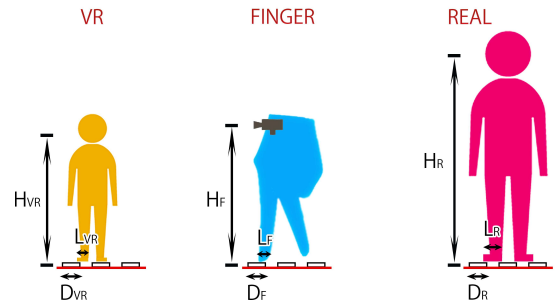}
  %
  %
  \caption{\label{fig:2}
           Geometry of an avatar, fingers, and a user's body.}
\end{figure}

When anthropomorphic finger motions are used, it is important to note that there are three models concerning users: the virtual avatar in the VEs, a user's finger and a user in the real world (Figure~\ref{fig:2}).
Each model has three parameters: the size of the model, $H$, the foot size, $L$, and the feature size of the texture, $D$.
The ratio of the foot size to the height of the virtual avatar and the user is expected to be the same.
\begin{equation}
{L_{VR}}=\frac{L_{R}}{H_R} \times {H_{VR}},
\end{equation}
where the user's height, $H_{R}$, which also represents the avatar's body size, and foot size, $L_{R}$, are known by the user from his/her experience, and the size of the avatar, $H_{VR}$, is provided by the first person view.
Therefore, the virtual avatar's foot size, $L_{VR}$ can be comprehended by the user.
Organizing our hypothesis in line with the Figure~\ref{fig:2}, users compare the ratio of the size of the contact area to the texture size of fingers in the real world and the virtual avatar.
\begin{eqnarray}
{D_{VR}}&=&\frac{L_{VR}}{L_F} \times {D_F}\\
\label{eq:eq3}
&=&\frac{L_{R}}{L_{F}} \times \frac{H_{VR}}{H_{R}} \times {D_{F}},
\end{eqnarray}
where the sense of touch gives the contact size of the finger, $L_{F}$.
Therefore, it is expected that there will be two methods for varying the interpretation of the tactile sensation, $D_{VR}$.
One method is to take a texture, $D_F$, the size of which is varied, and the second is to alter the avatar's body size, $H_{VR}$.
If this relationship is clarified, we can decide the texture to be displayed for the fingers based on the ground texture in the VEs.
We conducted three experiments to verify our hypothesis.
In experiment 1, illusory ownership while walking in the virtual world using an anthropomorphic finger motion interface was evaluated by means of a questionnaire.
In experiment 2, the avatar's body size was determined as a constant value and we investigated the effect of texture feature size, $D_F$, on the interpretation of the tactile sensation, $D_{VR}$.
In experiment 3, the effect of the foot size of the avatar on the interpretation of $D_{VR}$ was investigated.
%
\section{Experiments}
%
\subsection{System}
The participant was sitting on a chair and the distance between the participant and the display was 60 cm.
The participant reached out his/her arm to his/her for the front right.
A standing screen was positioned beside the hand to hide it from the participant's view.
The participant's palm and fingers were tracked by the Leap Motion Controller.
The Unity3D game engine displayed the graphical perspective that synchronized with the virtual camera on the back of the hand.
The virtual camera, which could freely move in space, had three translational degrees of freedom.
An artificial tactile board with individual raised portions was used to give users tactile sensation.
%
\begin{figure}[htb]
  \centering
  \includegraphics[width=.7\linewidth]{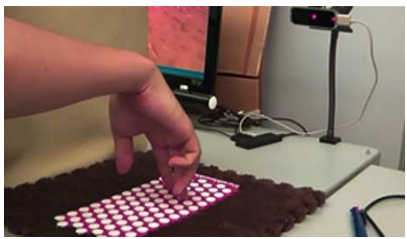}
  %
  %
  \caption{\label{fig:3}
           Sensing walking motions and tactile feedback.}
\end{figure}
%
For the experiments, 21 texture materials were 3D printed.
Each board had a pattern of circles, where the diameter of the circles ranged between 5 and 15 mm, increasing by 0.5 mm at every instance.
The diameter, height, and pitch of the circled pattern were proportionally scaled.
In the rest of this paper, the diameter of the circle is regarded as an index of texture feature size.
%
\begin{figure}[htb]
  \centering
  \includegraphics[width=.7\linewidth]{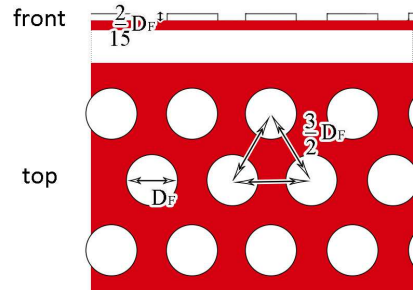}
  %
  %
  \caption{\label{fig:4}
           Artificial tactile board.}
\end{figure}
%
\subsection{Experiment 1}
\begin{figure}[htb]
  \centering
  \includegraphics[width=.99\linewidth]{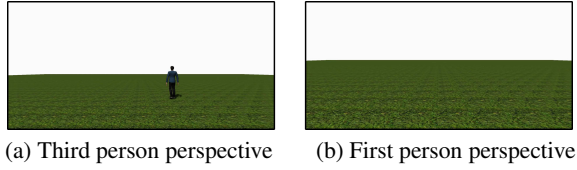}
  %
  \parbox[t]{1.\columnwidth}{\relax
  (a) Third person perspective\hspace{\fill}(b) First person perspective
           }
  \caption{\label{fig:5}
	First person and third person perspective.}
\end{figure}

The objective was to clarify that the synchrony between the first person perspective and the proprioceptive information together with the motor activity of the fingers are able to induce an illusion of ownership of the virtual legs.
Even if the target body part was invisible, we expected that the illusory ownership would be effected by imagining the avatar's legs from the first person perspective.
The participant's task was to "walk" back and forth and around on the planetable for one minute and then answer the questionnaire.
We experimentally compared a synchronous and asynchronous condition from the first and third person perspective.
Under the third person view condition, participants obtained the graphical perspective rendered from a fixed position behind the virtual avatar and observed the motions of the avatar's legs (Figure~\ref{fig:5}).
This situation is similar to that in the real world, where the users operate their fingers in front of them.
Following the previous studies \cite{sanchez2010virtual}\cite{kokkinara2014measuring}, we conducted experiments under synchronous and asynchronous conditions and compared the difference in the results thus obtained.
In the synchronous condition, the movement of the participant's fingers captured by the Leap Motion Controller determined the movements of the virtual avatar's leg motion and viewpoint.
In the asynchronous condition, the virtual avatar's movements were prerecorded and thus it moved independently of the movements of the participants's fingers.
%
Referring to the previous visuomotor studies, the illusion was measured with a self-report questionnaire adapted from Botvinick and Cohen's study\cite{botvinick1998rubber} and comprised five questions to which the participants responded on a 7-point Likert scale, where a score of 7 was described as “totally agree” and a score of 1 as “totally disagree” with the assertion.
The five questions were chosen on the basis of whether they could be reasonably adapted to the current study (Table~\ref{table:1}).
%
\begin{table}[htb]
\caption{Questionnaire}
\label{table:1}
\center
\includegraphics[width=1.0\linewidth]{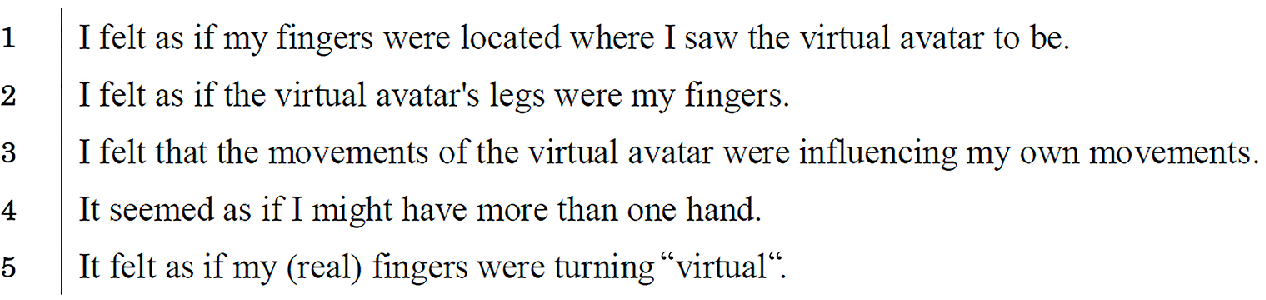}
\end{table}
The participants in this experiment were five males.
All were right-hand dominant according to their own judgement.
%
\subsection{Results}
\begin{table}[htb]
\caption{Left: first person perspective, Right: third person perspective}
\label{table:2}
\center
\includegraphics[width=1.0\linewidth]{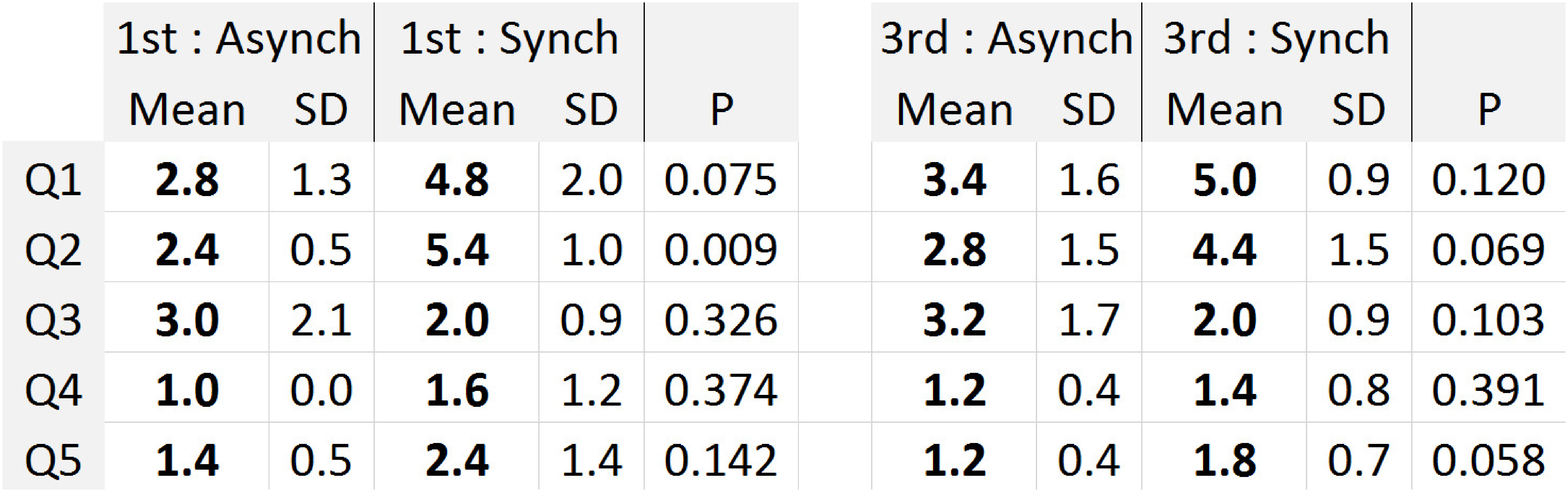}
\end{table}

To evaluate the hypothesis regarding the role of the synchronous visuo-motor effect, a dependent group t-test comparing the synchronous condition with the asynchronous condition for each view condition was conducted.
Table~\ref{table:2} shows the means, standard deviations, and the significance level for the difference in both view conditions.
In the third person view condition, in the responses to the questionnaire items addressing the illusion of ownership (Q1, Q2), those for Q2 had a significantly different mean (p < 0.1).
Responses to Q3 and Q4 did not have significantly different means.
Responses to Q5 had a significantly different mean.
As for the first person perspective, the responses to both the questionnaire items that addressed the illusion of ownership questions had significantly different means (Q1: p < 0.1, Q2: p< 0.05).
However, none of the responses to the remaining items had significantly different means.


The results we obtained were similar to those presented in \cite{sanchez2010virtual}.
Whether illusory ownership was generated should be determined by the responses to Q1 and Q2 for the following reasons.
Previous studies \cite{sanchez2010virtual}\cite{dummer2009movement} yielded inconsistent results for Q3 and Q4.
For Q5, both of the previous studies \cite{sanchez2010virtual}\cite{dummer2009movement} yielded significant differences.
The original rubber hand illusion\cite{botvinick1998rubber} did not yield significantly high scores for Q3 and Q4.
Comparing both view conditions, in the case of the first person perspective condition, the illusory feeling that was equivalent to the sense of ownership was strongly induced.
This experiment could not clarify that the feeling induced in the first person perspective condition was the same as the sense of ownership mentioned in the original rubber hand illusion paper \cite{botvinick1998rubber}, because the target body part in this experiment was invisible.
On the other hand, in the case of the third person perspective condition, Q1 did not yield a significant difference, although the overall results showed a trend similar to that found in previous studies.
This is partly because the position of the avatar was not the same as that of the participant's fingers from the participant's viewpoint.
The third person perspective condition is similar to the previous studies investigating visuomotor condition\cite{sanchez2010virtual}\cite{dummer2009movement}.
Therefore, the illusory ownership in the third person perspective condition will be clarified in further experiments with more participants.
An evaluation other than administering a questionnaire should be used.
In this experiment, a feeling that was equivalent to illusory ownership of the invisible, different body part was confirmed for the first time in the first person view condition.
Under this condition, the participants experienced the tactile sensation on the avatar's sole through their fingers.

In the rest of this study, the first person view system was adopted and two experiments explored the interpretation of tactile stimuli in this settings.

\subsection{Experiment 2}

\begin{figure}[htb]
  \centering
  \includegraphics[width=.99\linewidth]{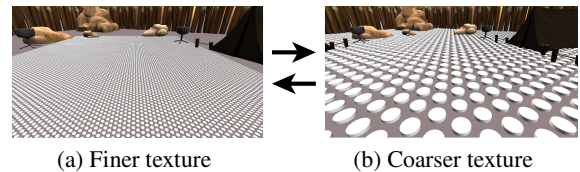}
  %

  \parbox[t]{1.\columnwidth}{\relax
  \hspace{18pt}(a) Finer texture\hspace{\fill}(b) Coarser texture\hspace{18pt}
           }
  \caption{\label{fig:6}
           Change of texture size in VEs.
           }
\end{figure}

The objective was to test whether the tactile perception, $D_{VR}$, was extended using anthropomorphic finger motions.
We used the following procedure in this experiment.
First, a tactile board was set under the participant's fingers. We call this the standard texture.
The participant was asked to walk on the board operating an avatar of standard size.
The standard size of the avatar was determined in comparison to the virtual objects around the avatar.
The participants' task in this experiment was to identify the ground texture in the VE that matched the tactile perception using key operation (Figure~\ref{fig:6}).
%
%
Let $D_{VR}$ be a diameter of each circle of the texture and $H_{VR}$ be the average height of the avatar's view, which is calculated based on the simulation.
The ratio of the diameter of the texture to the average height of the viewpoint is $D_{VR}/ H_{VR}$.
In order to relate the $D_{VR}$ to the value in the real world, the product of $D_{VR}/ H_{VR}$ should be multiplied by $H_R$, the height of the user in the real world.
Let $D_R$ be an interpreted texture size, and it should be a constant with regard to $D_{F}$. Let R be an enlargement ratio.
\begin{eqnarray}
D_R&=&\frac{H_R}{H_{VR}} \times D_{VR}\\
&=&\frac{L_{R}}{L_{F}} \times D_{F}
\end{eqnarray}
\begin{equation}
R=\frac{D_R}{D_{F}}=\frac{L_{R}}{L_{F}}
\end{equation}

Eight male graduate students took part in this experiment (mean age 22.75 years).
All were right-hand dominant according to their own judgement.
Nine tactile boards, the diameter of which ranged from 6 to 14 mm, were used.
Each participant tested all the boards in random order for two sets.

\subsubsection{Results}

\begin{figure}[htb]
  \centering
  \includegraphics[width=.99\linewidth]{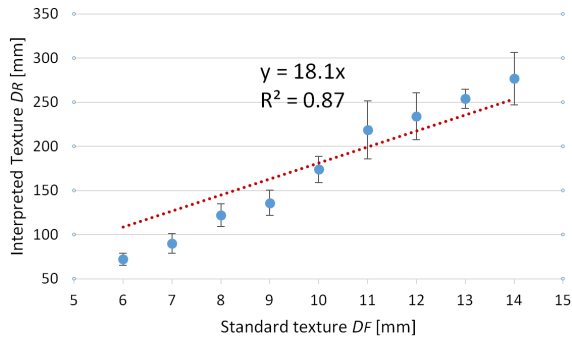}
  %
  %
  \caption{\label{fig:7}
           Relationship between the standard and interpreted texture.}
\end{figure}

Figure~\ref{fig:7} illustrates the relationship between the actual texture, which was presented as the standard texture $D_F$, and the interpreted texture $D_R$, namely, enlargement ratio R.
The comparison of the interpreted texture $D_R$ and the standard texture $D_F$ is plotted in blue.
The linear fits of the haptic judgement as a function of the comparison standard texture are represented by the red dotted line.
The approximately straight line is given by $y=18.1x$.
The enlargement ratio was approximately a constant and 18:1 in this experiment.
It is suggested that the texture under the avatar's feet was largely interpreted on the basis of the size of the tactile board in the real world.
%
%

The results show that the participants interpreted their haptic perception of the texture as that it was coarser.
Considering the length of the contact area on the fingertip is approximately 1.0 cm, and that of the human foot is from 23 to 30 cm, the enlargement ratio was smaller than expected.
However, shape of their contact area was different and would have an effect on the measured value.
If the length of the part vertical to the direction of the movement was an important factor in rescaling the haptic perception, the measured value would be reasonable.
The ratio was reduced when the standard texture was smaller.
The enlargement ratio differed among participants.
This is the reason why the standard error tended to increase with the increase in the standard texture.
The figure of the enlargement ratio is not important per se.
In contrast, it should be noted that the enlargement ratio of the perception of the participants is consistent.
It is suggested that each individual understood the correspondence relation among the three models in Figure~\ref{fig:2} in their own way.
It should be noted that the enlargement ratio, $R$, is not dependent on the height of the avatar, $H_{VR}$, but on the difference in the contact area (between fingers and feet).
The $H_{VR}$ should have an influence on the $D_{VR}$.

\subsection{Experiment 3}

\begin{figure}[htb]
  \centering
  \includegraphics[width=.99\linewidth]{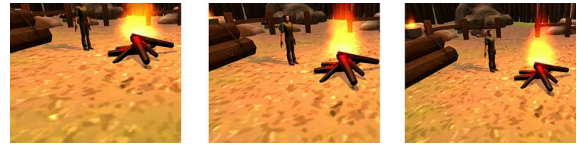}
  %
  \parbox[t]{1.\columnwidth}{\relax
         \begin{tabular}{rcl}(a) Two-thirds & \hspace{20pt}(b) Standard\hspace{10pt}    & (c) One and a half
  \end{tabular}
           }
  \caption{\label{fig:8}
           The avatar's body size was determined by the height of the perspective and the moving speed.}
\end{figure}

According to Eq. (\ref{eq:eq3}), varying the avatar's foot size, $L_{VR}$, would allow control of the interpretation of the tactile sensation, $D_{VR}$.
Therefore, we conducted an experiment to investigate the effect of the avatar's foot size, $L_{VR}$, on the tactile perception, $D_{VR}$.
In this experiment, the variant was not only the avatar's foot size, $L_{VR}$, but also the avatar's body size, $H_{VR}$.
This is because, in general, we are not conscious of foot size.
If the body size is varied, one would expect the foot size to be proportionally varied.
It was theoretically expected that the tactile perception, $D_{VR}$, would be proportional to the body size of the avatar, $H_{VR}$.
Although users did not know the body size in advance, they inferred it from the height of the perspective and the moving speed (Figure~\ref{fig:8}).

We used the following procedure in this experiment.
First, a tactile board was set under the participant's fingers.
We call this standard texture $D_F$.
The participant was asked to walk on the board operating an avatar of standard size, and imagine the feature texture size, $D_{VR}$.
Then, the body size was modified and participants were asked to select the texture material that invoked a similar perception regarding the texture feature size, $D_{VR}$ in the VEs.
This was called selected texture in this experiment.
We conducted this experiment under two conditions: the avatar's size was two-thirds (scale-down) or one and a half (scale-up) of the standard size.
It was theoretically expected that the tactile perception in the VEs would be proportional to the body size of the avatar.
If the subjects made a judgment based on the type of texture, regardless of avatar's size, the result would be close to that of the standard size.
On the other hand, if the interpretation was completely based on the avatar's size, the result would be close to the scaled value.
%
Nine male graduate students took part in this experiment (mean age 22.67 years).
All were right-hand dominant according to their own judgement.
They were unaware of the hypothesis under investigation.

\subsubsection{Results}

\begin{figure}[htb]
  \centering
  \includegraphics[width=.99\linewidth]{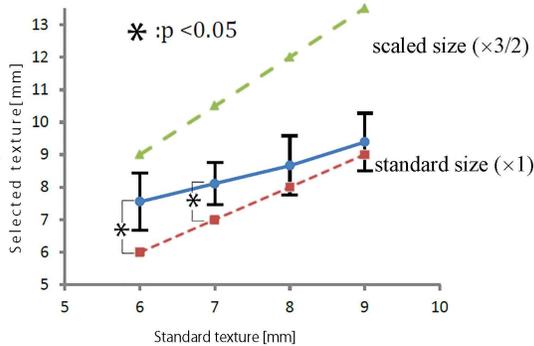}
  %
  %
  \caption{\label{fig:9}
           Perceived texture size in two-thirds size condition.}
\end{figure}

\begin{figure}[htb]
  \centering
  \includegraphics[width=.99\linewidth]{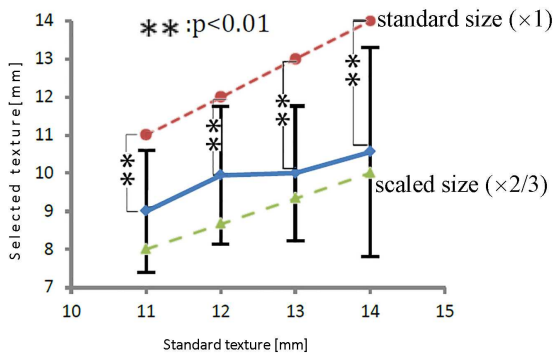}
  %
  %
  \caption{\label{fig:10}
           Perceived texture size in one and a half size condition.}
\end{figure}

Figures~\ref{fig:9} and \ref{fig:10} illustrate the results.
The horizontal axis indicates the standard texture when users operated the standard size avatar.
The vertical axis corresponds to the texture that was selected as producing the same size perception as the standard texture when operating different sized avatars.
If not affected by the avatar's size, the relation is plotted as the red line in the figures.
Conversely, if ideally affected, the relation is plotted as the green line.
The measured value is represented by the blue line.

%
Figure~\ref{fig:9} illustrates the relationship between texture feature sizes in the case of the avatar of the standard size and the avatar that was two-thirds of the standard size.
The participants chose the larger size material on average, and the difference was statistically significant in cases where the standard texture size was 6 and 7 mm (p < 0.05).
Figure~\ref{fig:10} presents the relationship between the results for an avatar of standard size and one of one and a half times the standard size.
In this case, smaller-sized materials were selected (p < 0.01).
It follows that the tactile size perception, $D_{VR}$, was magnified or reduced based on the size of the avatar's body, $H_{VR}$.

In the ideal case, when the interpretation is fully based on the avatar size and the avatar size has been correctly recognized by the participants, the tactile perception will be proportional to that of the avatar.
The experimental results show that the tactile perception was closer to ideal when using a large avatar than when using a small one.
A comparison of the results of experiments 2 and 3 suggested that an effect closer to the ideal was found in experiment 2.
Two main factors explain the difference.
First, immersion in the nonstandard size avatar is an unfamiliar experience.
Second, there was a difference in both the quality and quantity of the procedure.
In experiment 3, participants were asked to imagine the size perception of the standard texture using the standard size avatar and to select the texture that gave them the same size perception.
The task included an operation that was inverse to that in experiment 2.
It was a high level task that included storing the information of size perception of tactile sensation and extracting it.
Although there was no time limit in either experiment, it took the participants almost three times as long to select the texture in experiment 3 as in experiment 2.
This showed the high degree of difficulty of experiment 3.

\section{Discussion}

The questionnaire results in experiment 1 were not sufficient to argue whether the sense of ownership was generated in this study.
In the previous studies, a change in haptic perception was observed only when illusory ownership was induced.
Therefore, the results of experiments 2 and 3 support the inducement of ownership.
However, it should be noted that the causal connection between sense of ownership and haptic perception was not clarified in this paper.
This study separately showed that synchronous stimulation could cause a feeling that is equivalent to illusory ownership and the plasticity of haptic perception.
%
%
\section{Conclusion}

The purpose of this study was to show the possibility that fingers can be a replacement for legs in locomotion interfaces in terms of tactile perception.
First, a preliminary experiment was conducted to investigate the inducement of illusory ownership.
The results of a self-report questionnaire showed that the synchrony between the first person perspective and proprioceptive information together with motor activity of the fingers are able to induce an illusion that is equivalent to the sense of ownership of an avatar's invisible virtual legs.
Second, the plasticity of the tactile perception using an anthropomorphic finger motion interface was investigated.
The experimental results suggested the participants interpreted the tactile sensation on the basis of the difference in scale between fingers and legs.
The tactile size perception was proportional to the avatar's body (foot) size.
Therefore, in order to display the target tactile perception to users, we have only to control the virtual avatar's body (foot) size and the roughness of the tactile texture.


\bibliographystyle{eg-alpha-doi}

\bibliography{biblio}
\end{document}